\documentclass[
     final            
]{aipproc}

\layoutstyle{8x11single}

\begin{document}

\title{Constraints, limits and extensions\\ for nuclear
energy functionals}

\classification{\texttt{21.10.Re,21.30.Fe,21.60.Jz,21.65.Ef}}

\author{Gianluca Col\`o}{
address={Dipartimento di Fisica, Universit\`a degli Studi,
         and INFN, Sez. di Milano, via Celoria 16, 20133 Milano (Italy)}
}

\begin{abstract}
In the present contribution, we discuss the behavior of Skyrme
forces when they are employed to study both neutron stars and 
giant resonance states in $^{208}$Pb within the 
fully self-consistent Random Phase Approximation (RPA). We point
out that clear correlations exist between the results for the
isoscalar monopole and isovector dipole resonances (ISGMR and 
IVGDR), and definite quantities which characterize the
equation of state (EOS) of uniform matter. We propose that the
RPA results or, to some extent, the mentioned EOS parameters, 
are used as constraints when a force is fitted. This suggestion 
can be valid also when the fit of a more general
energy density functional is envisaged. We use our considerations 
to select a limited number of Skyrme forces (10) out of a large sample
of 78 interactions.  
\end{abstract}

\maketitle


\section{Introduction}

The quest for an accurate density functional for atomic nuclei lies at
the forefront of nuclear structure research. Nuclei are
strongly interacting many-body systems. Except for the lightest
among them, namely those having mass number $A$ smaller than 
$\approx$ 10-15 \cite{Wiringa}, trying to include explicitly all
correlations in the wavefunction is not feasible. One is obliged
to reduce the complexity of the wavefunction and to employ effective 
interactions $V_{\rm eff}$. Since more than three decades, one of the most
widely used approaches in nuclear physics is the so-called 
self-consistent mean-field (SCMF) approach. In it, starting 
from an effective Hamiltonian $H_{\rm eff}=T+V_{\rm eff}$ where
$T$ is the kinetic energy, one calculates the total energy $E$ as the 
expectation value of that Hamiltonian over the most
general Slater determinant $\vert\Phi\rangle$, i.e., over the 
most general one-body density $\rho$ which is compatible with 
the symmetries of the system under study. This defines an energy density
functional $\cal E$, that is, 
\begin{equation}\label{general_E}
\langle \Phi \vert H_{\rm eff} \vert \Phi \rangle = E[\rho] =
\int d_3r\ {\cal E}(\rho).
\end{equation} 
By minimizing the total energy $E[\rho]$, one can derive the nuclear 
ground-state. In the simplest case, that is, in a system which is not
superfluid, this is achieved by means of the Hartee-Fock (HF) equations.
Among effective interactions, the zero-range forces
of the Skyrme type or the finite-range Gogny interactions, are the 
most popular. Their parameters are fitted on a limited number of 
known properties (saturation of nuclear matter, energies and radii 
of few magic nuclei). The relativistic mean field (RMF) models 
share the same philosophy, that is, the number of parameters is 
comparable and they are fitted in a similar way. For a recent review
about the mean-field methods, their advantages and the obvious
limitations, one can refer to \cite{Bender}.

Recently, many authors have pointed out that the nuclear energy
functional can be more general and not necessarily obtained from an 
effective Hamiltonian. There are groups, around the world, 
who are working intensively with the aim of writing 
directly the
energy density functional with the most general structure 
compatible with symmetries, and of fitting the parameters at that 
level. The rationale behind this, is the guarantee that an exact
functional exists (provided by the Hohenberg-Kohn theorem). 
An alternative strategy consists in trying to derive the energy
functional from an underlying theory, whether Br\"uckner-Hartree-Fock
or Dirac-Br\"uckner.

While these attempts are certainly of paramount importance, they
cannot suceed without a physical guidance. So, there is still
work to be done on the existing functionals, albeit limited in their
structure and not derived from an underlying theory. One should
\begin{itemize}
\item remove the approximations, if any, which are still present in the 
calculations for the ground-state and the excited states within the 
SCMF implementations; 
\item provide a clear link between functional parameters and observables;
\item propose extensions of the existing functionals as much as this
appears necessary to account for measured observables.
\end{itemize}

The basic issue is to define the observables. We mentioned above 
those associated with the ground-state: the total energy and 
the density with the quantities that can be derived like radii, 
quadrupole moments etc. Their relevance is undeniable, and at
the same time they have been already 
much discussed (see, e.g., \cite{Pearson} 
for a discussion of functionals which are fitted to nuclear
masses on a large scale). Much less attention is usually
paid to excited states. However, there exist states, 
like the giant resonances, whose
properties carry general and relevant nuclear structure
information. Consequently, we focus
in this contribution on elucidating the links between  
the giant resonance properties and specific features of the
existing Skyrme functionals. We then exploit these links by
proposing a selection of Skyrme forces; our first screening 
is actually based on neutron star properties, following 
closely the work of Ref. \cite{Stone}.

Our discussion is based on the assumption of a relationship
between the parameters of a functional and the results
obtained from specific calculations of the excitation modes
(in the case under study, the giant resonances). If, as
recalled above, the HF equations provide the nuclear ground-state
by minimizing the total energy, the corresponding time-dependent
(TD) equations describe the oscillations around that 
minimum. In the limit of small amplitude, the TDHF equations
reduce to the equations of the so-called Random Phase 
Approximation (RPA). RPA is a suitable theory to describe the nuclear giant 
resonances (although it cannot account for their spreading width).
In self-consistent RPA, the residual interaction is derived
from the ground-state mean field. Therefore, the RPA results
depend only on the parameters of the effective Hamiltonian.
In this way, one is able to link these parameters with the 
giant resonance properties, or with specific EOS parameters as we discuss
below. Within phenomenological RPA, based e.g. on a 
Woods-Saxon mean field and a residual interaction
which is fitted {\em ad hoc}, it is impossible
to establish these links. 

Most of the existing nuclei are open-shell systems in which the
pairing correlations are active. In this case, the HF framework 
can be extended and one introduces the Hartree-Fock-Bogoliubov 
(HFB) one, in which a Slater determinant of independent quasiparticles
instead of independent particles, is assumed. The corrsponding 
linear response theory which describes the small oscillations is the
quasiparticle RPA (QRPA). In these approaches, the Skyrme force must
be supplemented by a pairing interaction which is not the focus
of our present discussion. We point out, however, in what follows, 
that pairing can have an effect also on high-lying states like 
giant resonances and impact our discussion of the relationships
between the parameters of a Skyrme functional and the EOS
quantities. 

\section{Sketch of the method to solve the (Q)RPA equations}

We have at our disposal a fully self-consistent scheme for both
RPA and QRPA. Some results from RPA have been first presented 
in Ref.~\cite{comex2}. Skyrme-RPA theory is well known since many 
years, especially in its matrix formulation. In our scheme, we
first solve the HF equations in coordinate space and calculate
the unoccupied states by using the resulting mean field and 
box boundary conditions (which means that the continuum is
discretized). We build a basis of particle-hole (p-h) configurations 
and we diagonalize the associated RPA matrix, by checking carefully 
that the basis is large enough so that our results are stable.
We should also mention that in our scheme there is no approximation 
in the residual interaction, in that all its terms are taken into 
account. In the calculations presented below, the box dimension
is typically between $\approx$ 3-4 times the size of the nucleus, and
unoccupied states up to $\approx$ 60-80 MeV are included 
in the model space. 

The extension of our model for open-shell nuclei, in the form
of a fully self-consistent QRPA based on HFB, has been presented
for the first time in \cite{Li}. The formalism is analogous to 
that of Ref. \cite{Terasaki}. The starting point is the solution of
the HFB equations in coordinate space. In this case, the basis
for QRPA is built using canonical states. This allows keeping
the equations reasonably simple, that is, the part of the QRPA
matrix associated with the residual interaction is the same as
in the case of BCS (whereas the part associated with the
unperturbed Hamiltonian has non-diagonal elements, since
the canonical states are not eigenstates of that Hamiltonian).
However, the price to be paid is that the 
canonical basis must be quite large (the energy cutoff 
is $\approx$ 150-200 MeV, at variance with the RPA case).

After solving the RPA or QRPA equations, we obtain the full
set of eigenvalues and eigenvectors and from them the strength
function $S(E)$ associated with, e.g., the IS monopole or IV dipole
operators. The moments of the strength function
are defined as $m_k=\int dE\ S(E) E^k$. The usual
centroid energy is defined as $E_0=m_1/m_0$. In certain cases,
and also in the discussion below, other definitions like
$E_{-1}=\sqrt{m_1/m_{-1}}$ (called sometimes the constrained
energy) are used. Obviously, all possible well-defined 
centroid energies coincide if the strength has a single,
symmetric peak. In cases like the IS quadrupole, where
there is a giant resonance but also a low-lying peak, 
centroid energies must be defined in a limited energy
interval.

We should point out that the quantity $E_{-1}=\sqrt{m_1/m_{-1}}$ 
could be obtained without resorting to a full QRPA calculation, 
at least in the nonrelativistic framework. In fact, $m_1$ can
be obtained from the Thouless theorem and $m_{-1}$ from the
dielectric theorem. These theorems, which have been known
for long time in the non-superfluid case, have recently been 
demonstrated in the case with pairing - that is, in the case
of self-consistent QRPA on top of HFB \cite{theorems}. 

\section{Reminder of the relevant EOS parameters}

In Eq. (\ref{general_E}) the energy density functional has
been written in a schematic, oversimplified form. In fact,  
for systems that are not symmetric in neutrons and protons,
the total energy must depend on both neutron and proton 
densities ($\rho_q$, where $q$ labels $n,p$). Moreover, in 
finite nuclei a local functional depends also on 
gradients of the densities, $\nabla\rho_q$, on kinetic energy 
densities, $\tau_q$, and on the so-called spin-orbit densities $J_q$
(for details, see Ref. \cite{Bender}). 

In uniform matter, only the dependence on spatial densities shows up. 
Instead of $\rho_n$ and $\rho_p$, one can employ as variables the 
total density $\rho$ and the {\em local} neutron-proton asymmetry,
$\delta \equiv \left( \rho_n-\rho_p \right) / \rho$ 
(this quantity should not be confused with the {\em global} asymmetry 
$(N-Z)/A$). In uniform asymmetric matter, we can further simplify 
${\cal E}(\rho,\delta)$ by making a Taylor expansion in $\delta$ and 
retaining only the quadratic term,
\begin{equation}\label{def_sym}
{\cal E}(\rho,\delta) \approx {\cal E}_0(\rho,\delta=0) +
{\cal E}_{\rm sym}(\rho) \delta^2 = {\cal E}_0(\rho,\delta=0) +
\rho S(\rho) \delta^2.
\end{equation}
It has been checked that the quartic term is negligible, for
Skyrme functionals, at the densities of interest for our 
discussion \cite{Trippa1,Trippa2}. 

The first term at the r.h.s. of Eq. (\ref{def_sym}) is the energy 
density of symmetric nuclear matter; for it, the minimum of the 
energy per particle $E/A={\cal E}/\rho$ is well known and used
when functionals are fitted. The curvature around this
minimum is simply related to the nuclear matter 
incompressibility, which reads
\begin{equation}\label{K}
K_\infty = 9\rho_0^2 {d^2\over d\rho^2}{{\cal E}_0\over
\rho}\vert_{\rho=\rho_0}.
\end{equation}

The second term at the r.h.s. of Eq. (\ref{def_sym}) defines 
the symmetry energy $S(\rho)$. This quantity, and in particular its 
density dependence, is presently 
much under debate and different contributions in the present 
volume deal with its determination, in keeping with its ubiquitous 
relevance in nuclear structure, heavy-ion reactions, and nuclear
astrophysics. 
The density dependence of the 
symmetry energy around the saturation density $\rho_0$ of symmetric 
nuclear matter can be expressed by means of 
\begin{equation}
S(\rho) = S(\rho_0)+S^\prime(\rho_0)(\rho-\rho_0)
+{1\over 2}S^{\prime\prime}(\rho-\rho_0)^2+\ldots 
\end{equation}
Usually, one defines $S^\prime(\rho_0) = L / 3\rho_0$ and 
$S^{\prime\prime} = K_{\rm sym} / 9\rho_0^2$; in fact, if 
$x={\rho-\rho_0\over 3\rho_0}$, $L$ and $K_{\rm sym}$ are
respectively ${dS\over dx}$ and ${d^2S\over dx^2}$.  

Starting from a different point of view, we can relate the ISGMR 
energy $E_{ISGMR}$ in a given nucleus to the 
so-called finite nucleus incompressibility 
$K_{\rm A}$ which has been introduced in Ref. \cite{Blaizot}, 
\begin{equation}
K_{\rm A}={m \langle r^2 \rangle_0 E^2_{ISGMR}\over \hbar^2}
\end{equation}
(where $m$ is the nucleon mass and $\langle r^2 \rangle_0$ 
is the ground-state expectation value). 
The interest of this quantity stems from the fact that if
we consider a local functional like Skyrme written for a
spherical system, and we calculate its second derivative
around the minimum using various simplifying hypotheses, 
the main one being the use of the so-called ``scaling model'', 
we can write $K_{\rm A}$ in a form analogous to that of the mass formula, 
namely \cite{Blaizot} 
\begin{equation}\label{likemass}
K_{\rm A} = K_\infty + K_{\rm surf} A^{-1/3} 
+ K_{\tau} \left( {N-Z\over A} \right)^2
+ K_{\rm Coul}{\rm Z^2\over A^{4/3}}.
\end{equation}
Moreover, we can show that 
\begin{equation}\label{ktau}
K_{\tau} = K_{\rm sym} + 3L 
- {27\rho_0^2 L\over K_\infty}{d^3{\cal E}\over d\rho^3}\vert_{\rho_0}.
\end{equation}
The last formula shows that a constraint on $K_\tau$ is reflected directly 
on the parameters which are associated with the density dependence of the 
symmetry energy, $L$ and $K_{\rm sym}$.

\section{Skyrme sets applied to neutron stars and to giant resonances}

The usual complain concerning Skyrme functionals is that there
exist too many different parametrizations, and it is true that probably
around 100 or more parameter sets have been introduced in the
literature. It should be stressed that not all of them are to be put 
on the same footing: whereas some sets have been used for long time 
after they have been first proposed, other sets are ``marginal'' in 
the sense that they have been fitted with very specific purposes and 
adopted only for one or few applications. We wish to propose
a strategy to limit the number of Skyrme parameter sets to be 
considered ``reasonable''; the hope is that the criteria we propose, 
eventually improved, can be used when the fitting of a universal
functional \cite{UNEDF} is envisaged. 

The strategy we propose is applied to an ensemble of 78 Skyrme forces
which is certainly large enough to demonstrate the effectiveness
of the method. It is impossible to recover the parameters 
of all Skyrme sets ever introduced, and this effort would be meaningless
since in principle many more sets can be produced. We start from the
work already done in Ref. \cite{Stone}: here, the authors consider 
as a starting point an ensemble of 87 forces which can be considered,
quoting their words, ``a representative sample of the Skyrme interactions
used in the nuclear physics applications''. These forces are reported 
in Table I of \cite{Stone} and the reader can find there the original
references, that we do not report here for the sake of brevity. 
Our starting sample is rather similar: we exclude, compare to  \cite{Stone} the 
sets SLy0, SLy1, SLy2, SLy3, SLy8, SLy9 (they are unpublished), 
SLy6, SLy7, SLy10 (they include the two-body center-of-mass correction), 
SkI1, SkI4, SkI6, SkO (they do not lead easily to convergent results), and 
we add the four forces Ska \cite{Kohler}, SK255, SK272 \cite{Shlomo} 
and LNS \cite{Cao}. 

We apply the following criteria:
\begin{enumerate}
\item We select, among the 78 forces, those which have an overall
satisfactory behavior as far as the density dependence of the symmetry 
energy in the range 0.1$\le \rho \le$0.6 fm$^{-3}$ is concerned 
(we follow closely Ref. \cite{Stone} for this point).
\item For these forces we calculate the IVGDR in $^{208}$Pb and we make 
a further selection, by demanding that the forces
reproduce the experimental value $E_{-1}$ = 13.46 MeV \cite{Dietrich} 
within $\pm$ 1 MeV. 
\item Finally, we also demand that the selected interactions reproduce 
the experimental value of $E_{-1}$ = 14.17 MeV for the ISGMR in 
$^{208}$Pb \cite{Youngblood}, with the same accuracy of $\pm$ 1 MeV. 
\end{enumerate}

The Skyrme sets which have ``survived'' this kind of selection will be
listed at the end of our discussion. 

\begin{figure}
\includegraphics[height=.35\textheight]{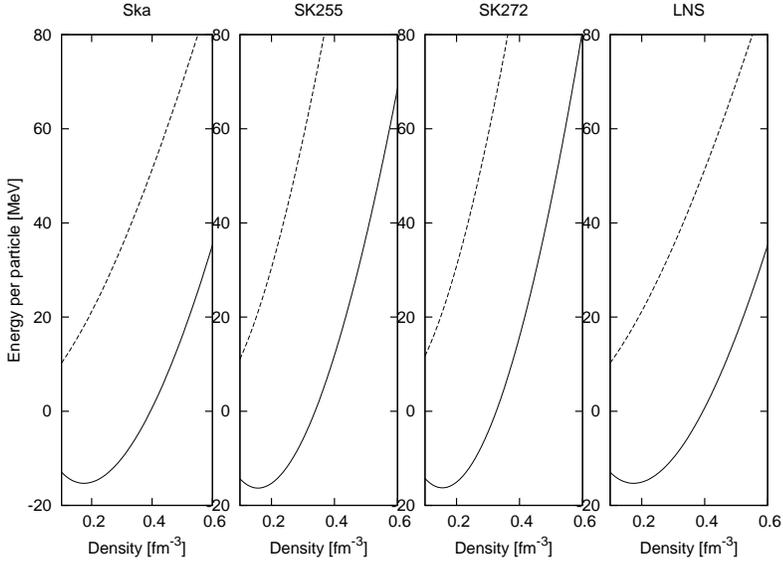}
\caption{For the four forces which label the panels (they are Skyrme sets
which have been not studied in Ref. \cite{Stone}), we display the energy
per particle in symmetric nuclear matter (full line) and in pure neutron
matter (dashed line). Because of the fact that these quantities have
a pronounced increase as a function of the density, and that the dashed 
curve never falls below the other one, these forces belong
to the group I defined in \cite{Stone} and lead to qualitatively correct
neutron star properties.}
\label{figure_esym}
\end{figure}

\subsection{Neutron stars and the overall behavior of the symmetry
energy}

In Ref. \cite{Stone} it has been tested whether Skyrme forces 
predict plausible neutron stars properties. The 
Tolman-Oppenheimer-Volkov (TOV) equation \cite{originalTOV,textbookTOV} 
has been
solved, coupled with the Skyrme EOS (supplemented by appropriate 
corrections for low densities), and the mass-radius relationship
associated with a given parameter set has been given. The result
is that Skyrme sets can be divided in three groups. Only the
sets of group I reproduce the expected qualitative 
relationship: this has been found to be strictly related with
the fact that the energy per particle increase quickly, as a function
of density, both in symmetric nuclear matter and in pure neutron matter, 
with the latter being always characterized by a larger energy with 
respect to the former. We have checked the performance of the
four sets that we have decided to add to the starting sample compared
with \cite{Stone}: in Fig. \ref{figure_esym} we have displayed the
mentioned quantities, so that it is clear that these interactions
obey the conditions which are sufficient to be included in the group I 
of the satisfactory parameter sets. Since the energy in pure neutron 
matter equals the energy in symmetric nuclear matter plus $S$, we
can say that in the proposed method of selection the overall 
behavior of the symmetry energy plays a key role.

In conclusion, after the selection we end up with 18 forces (Gs, Rs, 
SGI, SLy230a, SLy4, SLy5, SV, SkI2, SkI3, SkI5, SkMP, SkO$^\prime$, 
SkT4, SkT5, Ska, SK255, SK272, LNS). 

\begin{figure}
\includegraphics[height=.35\textheight]{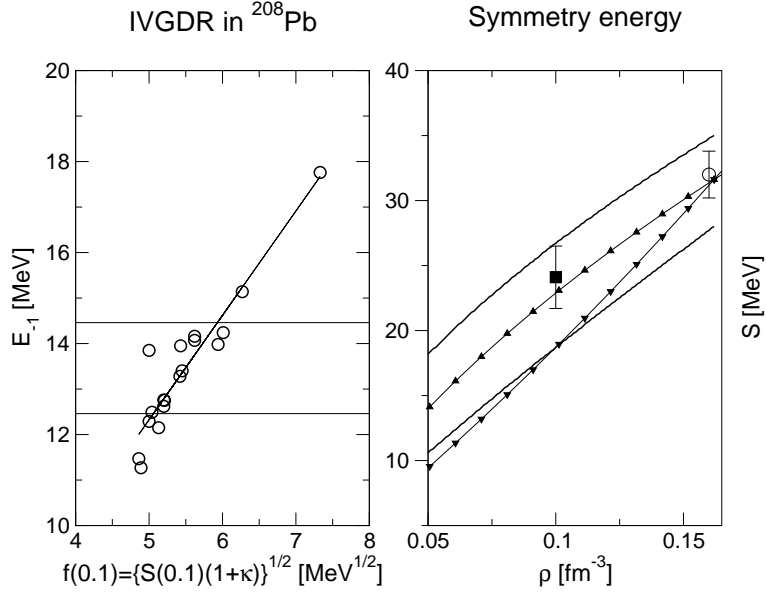}
\caption{In the left panel, the dipole energy is displayed as a function
of the quantity $f(0.1)$ defined in the text. The thin line is a linear
fit while the horizontal full lines correspond to the experimental
energy $\pm$ 1 MeV. In the right panel, the constraint on $S(0.1)$ 
extracted from the dipole is displayed with a black square and an associated
error bar. The open circle (also with error bar) shows the constraint
from Ref. \cite{Klimkiewicz}. 
The two thick lines, and the two lines which join small triangles, 
are bounds (upper and lower) for the symmetry energy $S(\rho)$ coming
from the studies respectively of Refs. \cite{Tsang} and \cite{BaoAnLi}.}
\label{figure_dipole}
\end{figure}

\subsection{The IVGDR and the related constraint on the symmetry energy}

With the 18 selected forces we have calculated the energy of
the IVGDR in $^{208}$Pb. The results are displayed in the left panel of
Fig. \ref{figure_dipole}, where the energy $E_{-1}$ (cf. above)
appears on the y-axis. On the x-axis the quantity
\begin{equation}
f(0.1) \equiv \sqrt{S(0.1)(1+\kappa)}
\end{equation}
is shown, where $S(0.1)$ is the symmetry energy evaluated
at $\rho$=0.1 fm$^{-3}$ and $\kappa$ is the so-called enhancement
factor of the IV dipole sum rule (with respect to the classical
Thomas-Reiche-Kuhn sum rule). The quantity $f(0.1)$ has been
already defined in \cite{Trippa1}, where physical arguments have
been provided to justify a correlation between $f(0.1)$ and 
the dipole energy $E_{-1}$. We do not repeat these arguments here.
We notice that the correlation is visible in the left panel of 
Fig. \ref{figure_dipole}, 
as testified by the thin line which corresponds to a linear fit. 
The two horizontal
lines correspond to the experimental dipole energy $\pm$ 1 MeV: 
the results associated with 12 forces (Gs, Rs, SGI, SLy230a, 
SLy4, SLy5, SkI3, SkMP, SkO$^\prime$, SK255, SK272 and LNS) lie 
within those lines, and these
forces are selected for further considerations. 

In Ref. \cite{Trippa1} the correlation between the dipole energy
and $f(0.1)$ has been used to extract a constraint on this latter
quantity. Unfortunately, $f$ contains at the same time the symmetry
energy as well as $\kappa$ and we miss an unambiguous experimental 
determination of the dipole enhancement factor. If one introduce 
an acceptable range for $\kappa$, between 0.18 and 0.26, $S(0.1)$
is constrained in the interval 24.1$\pm$0.8 MeV. However, we have
verified that in the ensemble of forces used in the present work (which
includes some forces that were not considered in \cite{Trippa1}),
there are some which do reproduce the experimental IVGDR energy 
having $S(0.1)$ outside the range of 24.1$\pm$0.8 MeV. To account
for this, we have displayed in the right panel of Fig. \ref{figure_dipole} 
the point corresponding to the dipole constraint with a larger error
bar, numerically equal to $\pm$ 3$\sigma$ (that is, 24.1$\pm$2.4 MeV).

In the same panel, results from other groups are reported. 
In Ref. \cite{Klimkiewicz} the data on the Pygmy Dipole Resonance (PDR)
obtained at GSI, Darmstadt has been compared with RMF calculations
and eventually a range of values for $J$ has been extracted, which
is shown in the panel with an open circle with its associated error
bar ($J$=32$\pm$1.8 MeV). In Ref. \cite{Danielewicz}, a throughout analysis
of nuclear surface symmetry energies has been carried out. This analysis led
to $J$=32.5$\pm$1 MeV which is compatible with the previous value.
We do not display the corresponding point in the figure, just to
keep it reasonably clear and readable. 

We also like to discuss, very briefly, the comparison with the 
constraints coming from studies of heavy-ion collisions. Data of 
isospin diffusion following the $^{112}$Sn-$^{124}$Sn reaction, 
have been anayzed using transport models, in particular the Improved 
Quantum Molecular Dynamics model (I) in \cite{Tsang} and the IBUU04 version 
of the Boltzmann-Uehling-Uhlenbeck (BUU) model (II) in \cite{BaoAnLi}. 
We are not in the position to discuss merits and pitfalls of these
models (for which we confer the reader to the original references).
However, for each model we draw two curves which correspond to 
acceptable upper and lower limits. These are the full thick lines 
in the case of model (I) and the thin lines joining the triangles
in the case of model (II). 

In principle, other observables can be very effective to constrain the
symmetry energy and its density dependence. We mention the sum rules 
of charge-exchange excitations \cite{Sagawa}, which have been
measured yet with unsufficient precision, and the neutron radius
of $^{208}$Pb \cite{Piekarewicz}, which should be very accurately
determined by the PREX experiment. 

\begin{figure}
\includegraphics[height=.35\textheight]{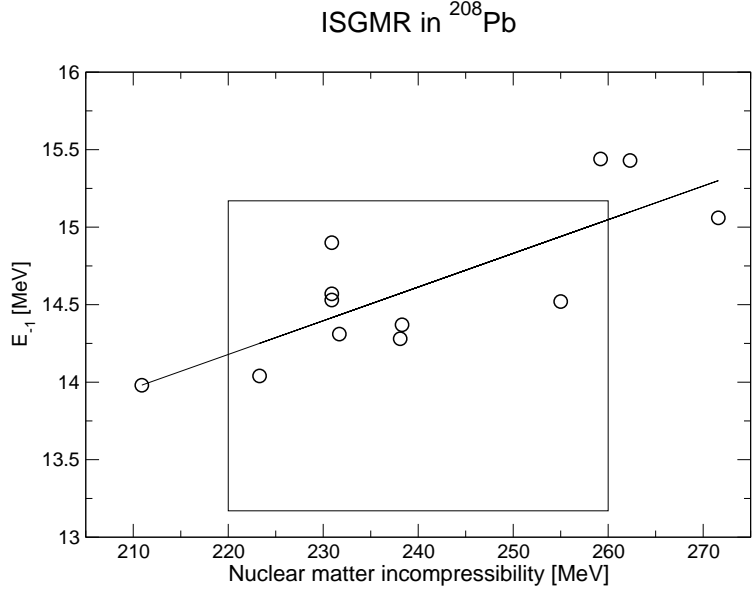}
\caption{The monopole energy is displayed as a function
of the nuclear matter incompressibility $K_\infty$ (cf. Eq. 
(\ref{K})). The box defines a region which corresponds to
$K_\infty$=240$\pm$20 MeV \cite{monopole_rev}, and to a 
monopole energy which is within $\pm$ 1 MeV with respect
to the experimental value. See the discussion in the main text.}
\label{figure_monopole}
\end{figure}

\subsection{The ISGMR and the constraint on the nuclear incompressibility}

With the 12 forces mentioned in the previous subsection, we have
calculated the ISGMR in $^{208}$Pb. The result for the energy $E_{-1}$
is compared with the experimental finding, namely 14.17 MeV \cite{Youngblood}. 
Almost all the forces that have been selected through the calculation of
the IVGDR, reproduce the ISGMR energy $\pm$ 1 MeV. Only SGI and SkI3
must be rejected, and we end up with a set of 10 forces, that is, 
Gs, Rs, SLy230a, SLy4, SLy5, SkMP, SkO$^\prime$, SK255, SK272 and LNS. 

The results are shown in Fig. \ref{figure_monopole}, as a function of
the associated value of the nuclear matter incompressibility 
$K_\infty$ (cf. Eq. (\ref{K})). 
The thin line corresponds to a linear fit. The open box defines a region 
which corresponds to $K_\infty$=240$\pm$20 MeV, and to a 
monopole energy which is within $\pm$ 1 MeV with respect to the experimental 
value. The preferred range given by $K_\infty$=240$\pm$20 MeV has been 
extensively discussed 
in \cite{monopole_rev}, and it actually comes from an analysis which 
includes not only Skyrme forces but a careful confrontation with results
obtained with Gogny interactions and RMF parametrizations as well, so that 
we believe its validity is rather general.
Also in Fig. \ref{figure_monopole}, almost all the forces which reproduce
reasonably the monopole energy, have an associated value of the
incompressibility in that range (there are only two exceptions). 

\begin{figure}
\includegraphics[height=.30\textheight]{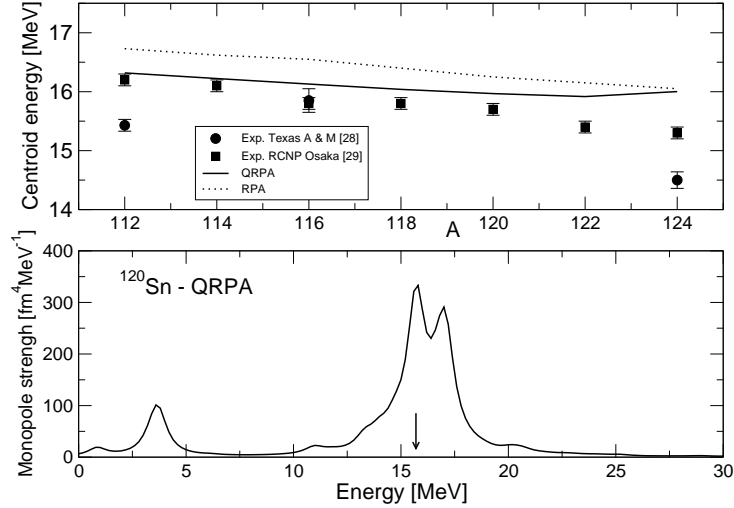}
\caption{In the upper panel, the experimental results for the
monopole centroid energies from \cite{Youngblood2004} and 
\cite{Li2007} are displayed. They are compared with the RPA and 
QRPA results of Ref. \cite{Li}. In the lower panel, the QRPA 
strength distribution is shown in detail for the case of 
$^{120}$Sn. The discrete QRPA results have been smeared out with
Lorentzians having 1 MeV width, only for illustrative purposes. 
The arrow indcates the position of the experimental centroid from
the upper panel.}
\label{figure_qrpa}
\end{figure}

\subsection{Remarks on the ISGMR in open-shell isotopes}

As a side remark, albeit quite important, we discuss in this subsection
the ISGMR in the Sn isotopes. In Ref. \cite{Piekarewicz}, the 
question has been raised ``why is tin so
soft'' or, in other words, why do theoretical models (with a value of
the incompressibility $K_\infty$ in the quoted range 
$K_\infty$=240$\pm$20 MeV) that 
reproduce the values of the ISGMR energy in $^{208}$Pb as well
as in $^{90}$Zr, tend to overestimate this energy in Sn isotopes ?
These are semi-magic nuclei and neutrons are superfluid. Answering to
the question above, implies among the rest a serious assessment of
the effect of pairing correlations on the ISGMR.

This has motivated the work of Ref. \cite{Li}, where a fully self-consistent
QRPA based on HFB has been applied to the study of the monopole strength
distribution in the Sn isotopes. The Skyrme force has been supplemented
with an effective, zero-range, density-dependent pairing force. Three
kinds of pairing forces have been tested, namely volume, surface, and
mixed pairing forces. Writing the pairing force as 
\begin{equation}
V_{\rm pair}(\textbf{r}_{1},\textbf{r}_{2})=V_{0}\left[1-\eta
\left(\frac{\rho(\frac{\textbf{r}_{1}+\textbf{r}_{2}}{2})}{\rho_{0}}
\right) \right]\delta(\textbf{r}_{1}-\textbf{r}_{2}),
\end{equation}
the three kinds of pairing correspond respectively to $\eta$=0, 1 and 1/2. 
$\rho_{0}$ is fixed at 0.16 fm$^{-3}$ and the values of $V_{0}$ (for every
kind of pairing but also for every Skyrme parameter set employed) are fixed 
by fitting the empirical value of the mean neutron gap of $^{120}$Sn. Pairing
is treated consistently in HFB and QRPA, and also the (small) pairing 
rearrangement terms have been analyzed. The reader can consult \cite{Li} 
for further details. 

Some of the main results of that investigation are in Fig. \ref{figure_qrpa}.
Looking at the upper panel, one can notice a systematic shift 
downwards of the QRPA results with respect to RPA. This shift is due
of the attractive monopole pairing matrix elements. It is not
constant along the isotope chain, but it tends to decrease with 
increasing $N-Z$: this effect has been explained in \cite{Li} as
a consequence of the level occupancies. In the
lower panel of Fig. \ref{figure_qrpa} we show the QRPA monopole
strength distribution in a typical case. 

This particular set of
results corresponds to the force SkM$^*$ plus surface pairing.
Since SkM$^*$ has an associated value of $K_\infty$ given by
217 MeV, looking at $^{112-120}$Sn one would conclude that
the discrepancy between the values of the nuclear matter
incompressibility extracted from the Sn isotopes and from 
$^{208}$Pb differ by about 10\%. Of course, this points 
to our still incomplete understanding of the details of the 
nuclear effective functionals - but the puzzle would be greater
if the pairing contribution had been overlooked. 

Still two considerations are in order. Firstly, in 
the two cases of $^{112}$Sn and $^{124}$Sn, the results of
the two experimental groups disagree quite seriously, but
understanding the reasons of this discrepancy is in 
progress \cite{Garg}. Secondly, the main motivation at the
basis of the experiment reported in \cite{Li2007} was the
extraction of the parameter $K_\tau$ defined
by Eq. (\ref{likemass}): in fact, the trend of the experimental 
data, plus the simplifying assumption which consists 
in setting $K_{\rm surf}\approx -K_\infty$, have allowed
extracting $K_\tau$ from the data. The simplifying assumption
is consistent with simple, geometrical arguments (the
same which apply to the liquid-drop mass formula). 
The coefficient $K_{\rm Coul}$ can be calculated. 
In this way, a value of $K_\tau$ equal to 550 $\pm$ 100 MeV
has been deduced.   

\begin{figure}
\includegraphics[height=.35\textheight]{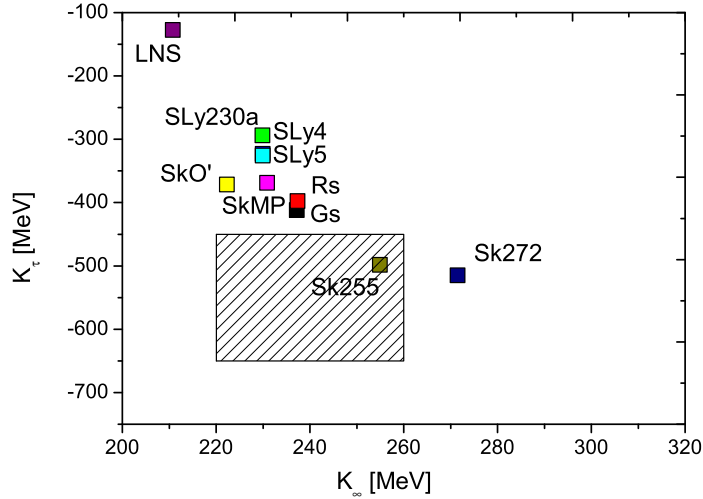}
\caption{Values of $K_\infty$ and $K_\tau$ associated with the
10 Skyrme forces which have been selected. The relevance of the
shaded box is discussed in the main text.}
\label{figure_ktau}
\end{figure}

Based on this result, one would be tempted to set another
constraint on the Skyrme forces: that is, to select
those with the empirical value of $K_\tau$. However, 
a strong warning is appropriate here. Comparing $K_\tau$ 
from the data with the result of Eq. (\ref{ktau}) neglects 
the fact this latter equation does not include any 
surface-symmetry contribution. We have plotted in 
Fig. \ref{figure_ktau} the values of $K_\infty$ and $K_\tau$
of the 10 selected forces which have been listed at the
end of the previous subsection. The shaded box defines the
intersection of the constraints on the quantities. Whereas 
the one on $K_\infty$ has been claimed to be robust, the
one on $K_\tau$ suffers from the mentioned drawback. 
There is a clear tendency of Skyrme forces to predict
values of $K_\tau$ which are smaller (in absolute value)
than 450 MeV. This remains true if a larger sample of
Skyrme parametrizations is considered. 

\section{Conclusions}

While part of the nuclear structure community is striving
to construct a universal, accurate energy density functional,
many calculations are still performed with e.g. Skyrme forces
which should probably be rejected. In fact, there is not
consensus on which properties should be necessarily
reproduced by a mean-field calculation with an effective
force, and what pitfalls should be tolerated. 

It is clear that many Skyrme interactions have been built
with an eye on very specific applications and should not
be used systematically. In this paper, we focus on the
performance of Skyrme forces when they are applied to
the study of giant resonances, in particular the ISGMR and
IVGDR. We do not calculate these modes by using a huge ensemble of
Skyrme parameter sets, but we use for the purpose of screening
the results of Ref. \cite{Stone}, that is, we demand first 
that the overall behavior of the energy per particle 
both in symmetric uniform matter and in neutron matter are
reasonable in the sense defined in the quoted work.

The results for the ISGMR and IVGDR in $^{208}$Pb are claimed
to be a valid constrain to be imposed on existing forces as
well as on envisaged new functionals. This in keeping with
the fact that we have shown that the constraints can be
translated into conditions on physical parameters which
characterize the nuclear EOS like the nuclear matter
incompressibility and the symmetry energy at sub-saturation
density. We have also demonstrated that
it is not straighforward to extend trivially the considerations
made for e.g. $^{208}$Pb, to the case of open-shell nuclei.
In the case of ISGMR, in particular, we have elucidated the
role played by the pairing correlations.

The ISGQR could be considered as well as an input for our
considerations, whereas in the case of spin and spin-isospin modes
probably extensions of the effective forces should be
envisaged (besides other reasons, to avoid instabilities of
uniform matter in the spin and spin-isospin channels 
\cite{Margueron}). 


\begin{theacknowledgments}
Many of the results reported here have been obtained through collaborations
with colleagues and students. In particular, the author would like to
thank L. Capelli, J. Li, J. Meng, L. Trippa, E. Vigezzi. Discusssions with 
U. Garg about the data and the analysis of Ref. \cite{Li2007} are gratefully
acknowledged. Thanks are also due to B. Tsang and B.A. Li for providing 
the author with the data displayed in Fig. \ref{figure_dipole}, and for 
clarifications about the issue of the symmetry energy extracted from the
study of heavy-ion collisions. The authors expresses special thanks to
P. Danielewicz for warning against a strict comparison of $K_\tau$ from
data and from Eq. (\ref{ktau}). 
\end{theacknowledgments}



\end{document}